\documentclass[aps,twocolumn,superscriptaddress,showpacs,showkeys,amsmath,amssymb]{revtex4}

\usepackage{epsfig,amsmath,amssymb,color}
\begin{document}

\title{Thermodynamic properties of Ba$_2$CoSi$_2$O$_6$Cl$_2$ in strong magnetic field:
       \protect\\
       Realization of flat-band physics in a highly frustrated quantum magnet}

\author{Johannes Richter}
\affiliation{Institut f\"{u}r theoretische Physik,
          Otto-von-Guericke-Universit\"{a}t Magdeburg,
          P.O. Box 4120, 39016 Magdeburg, Germany}

\author{Olesia Krupnitska}
\affiliation{Institute for Condensed Matter Physics,
          National Academy of Sciences of Ukraine,
          Svientsitskii Street 1, 79011 L'viv, Ukraine}

\author{Vasyl Baliha}
\affiliation{Institute for Condensed Matter Physics,
          National Academy of Sciences of Ukraine,
          Svientsitskii Street 1, 79011 L'viv, Ukraine}

\author{Taras Krokhmalskii}
\affiliation{Institute for Condensed Matter Physics,
          National Academy of Sciences of Ukraine,
          Svientsitskii Street 1, 79011 L'viv, Ukraine}
\affiliation{Department for Theoretical Physics,
          Ivan Franko National University of L'viv,
          Drahomanov Street 12, 79005 L'viv, Ukraine}

\author{Oleg Derzhko}
\affiliation{Institute for Condensed Matter Physics,
          National Academy of Sciences of Ukraine,
          Svientsitskii Street 1, 79011 L'viv, Ukraine}
\affiliation{Department for Theoretical Physics,
          Ivan Franko National University of L'viv,
          Drahomanov Street 12, 79005 L'viv, Ukraine}
\affiliation{Abdus Salam International Centre for Theoretical Physics,
          Strada Costiera 11, 34151 Trieste, Italy}

\date{\today}



\begin{abstract}
The search for flat-band solid-state realizations is a crucial issue 
to verify or to challenge theoretical  predictions for quantum many-body flat-band systems. 
For frustrated quantum magnets flat bands lead to various unconventional properties
related to the existence of localized many-magnon states. 
The recently synthesized magnetic compound Ba$_2$CoSi$_2$O$_6$Cl$_2$
seems to be an almost perfect candidate to observe these features in experiments.
We develop a theory for Ba$_2$CoSi$_2$O$_6$Cl$_2$ by adapting the localized-magnon concept to this compound.
We first show that our theory describes the known experimental facts 
and then we propose new experimental studies 
to detect a field-driven phase transition related to a Wigner-crystal-like ordering of localized magnons at low temperatures.
\end{abstract}

\maketitle

\section{Introduction}

Destructive interference in quantum mechanics can lead to a disorder-free localization of
particles.
In the one-particle energy spectrum this is related to the existence of flat bands.
Such a single-particle flat band can substantially influence the many-body physics of interacting quantum systems.
For two-dimensional (2D) electronic systems in a magnetic field this scenario may lead to so-called Aharonov-Bohm cages \cite{Vidal2000,Moller2012} 
as well as to the celebrated fractional quantum Hall effect \cite{Tsui1982}.
In correlated-electron systems with a flat band the interaction energy may dominate over the kinetic energy, 
thus a flat band may lead to ferromagnetic instability in the Hubbard model.
Remarkably,  
here the existence of a flat band allows for exact results in this highly non-trivial many-body system \cite{Mielke1991,Tasaki1992,Mielke1993,Derzhko2010a}.

Nowadays, flat-band (FB) physics is extensively discussed in the recent literature with a special focus on topological FB models, 
and many interesting phenomena related to flat bands have been observed, 
see, e.g., \cite{Parameswaran2013,Bergholtz2013,Leykam2013,Huber2010,Derzhko2015} and references therein.
Realizations of FB systems can be achieved, 
e.g., 
with cold atoms in optical lattices \cite{Jo2012,Struck2012} and photonic lattices \cite{Vicencio2015,Mukherjee2015,Baboux2016}.
On the other hand, solid-state realizations of ideal FB systems are notoriously rare, 
since a strictly flat band requires a perfect FB geometry providing immaculate Hamiltonian parameters.

Among the numerous FB systems, 
the highly frustrated quantum antiferromagnets (AFMs) play a particular role in solid-state physics.
These FB spin systems exhibit several prominent features in high magnetic fields,
such as a plateau and a subsequent magnetization jump at the saturation field \cite{Schnack2001,Nishimoto2013}, 
a magnetic-field driven spin-Peierls instability \cite{Richter2004},
a finite residual entropy at the saturation field \cite{Derzhko2004,Zhitomirsky2004,Derzhko2006},
and an unconventional low-temperature thermodynamics \cite{Zhitomirsky2004,Derzhko2006,Richter2006,Derzhko2010b}.
These unconventional features are related to the existence of a huge manifold of exactly known many-body low-lying eigenstates 
(the so-called independent localized many-magnon states)
which 
(i) allow an exact description of the low-energy physics 
and 
(ii) the calculation of the low-temperature thermodynamic properties
by mapping of the localized many-magnon states of the initial quantum spin system onto classical lattice-gas models of hard-core objects, 
for a review, see, e.g., \cite{Derzhko2015,Derzhko2007}.

\begin{figure}
\begin{center}
\includegraphics[clip=on,width=45mm,angle=0]{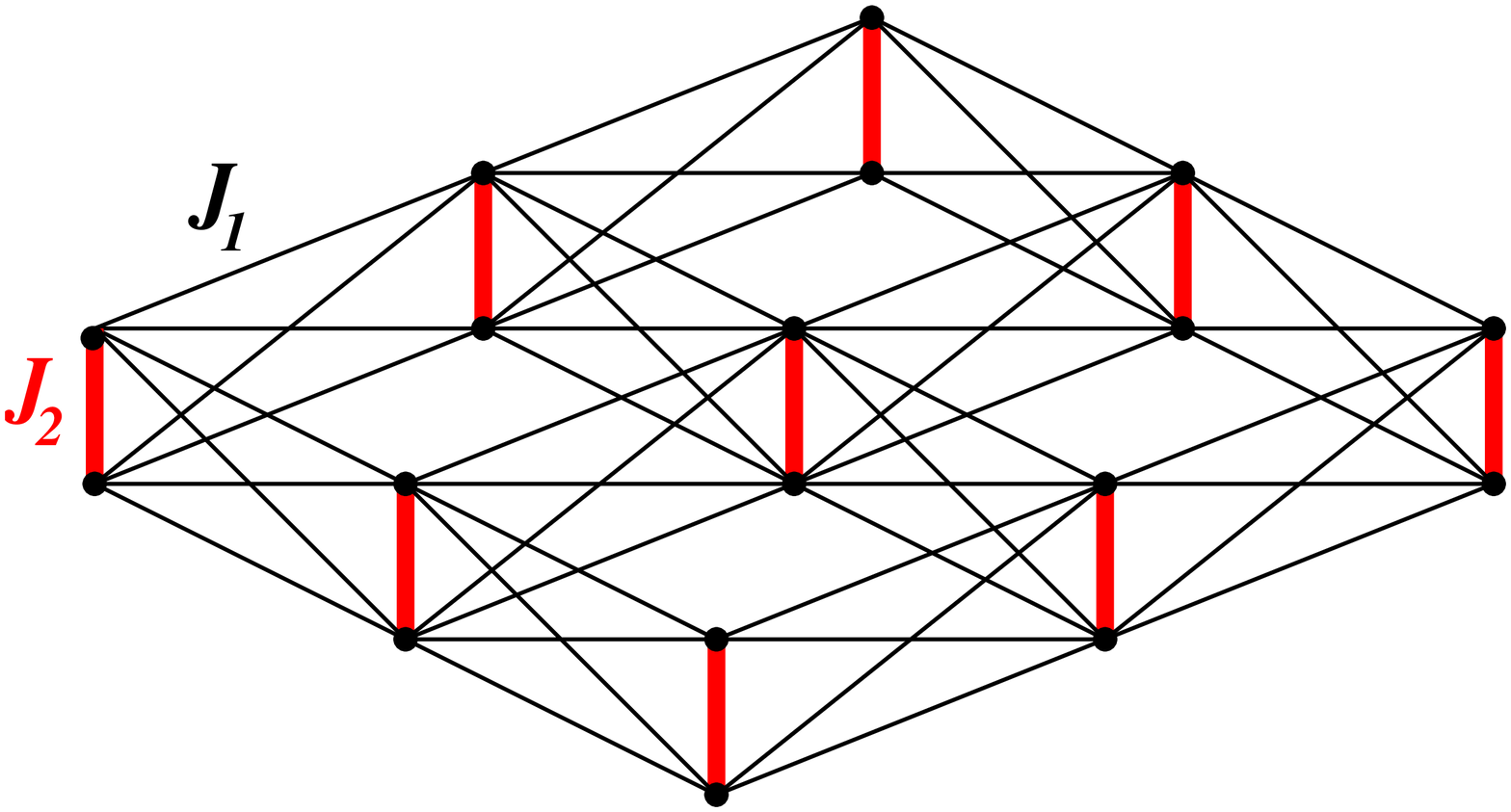}
\includegraphics[clip=on,width=60mm,angle=0]{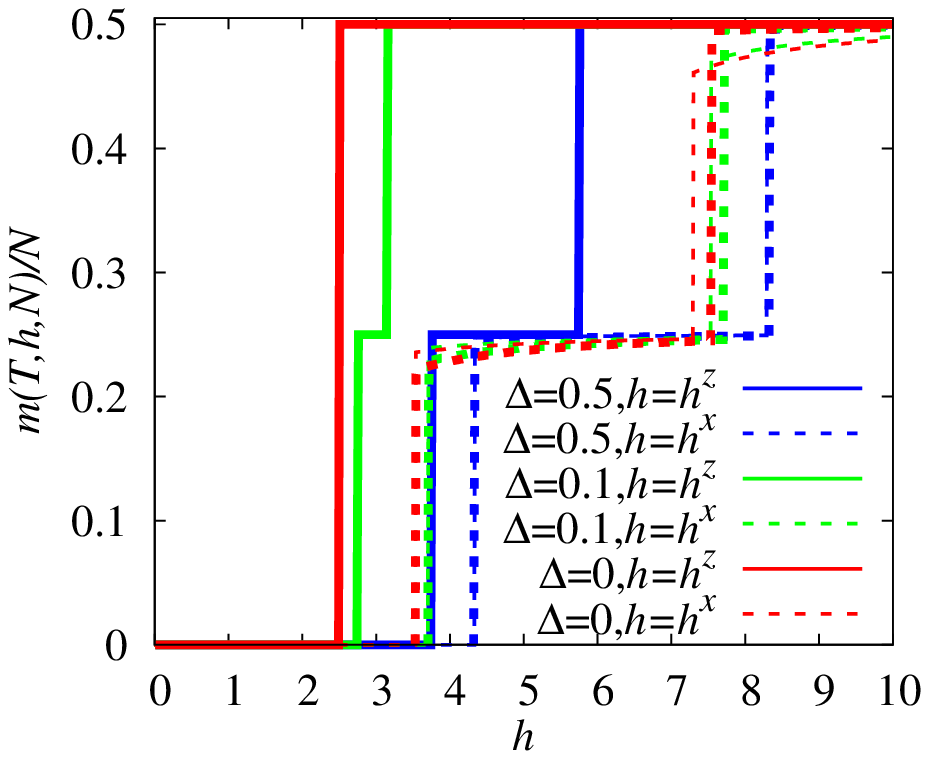}
\caption{(Top) Frustrated square-lattice bilayer.
Red bonds: intradimer bonds $J_2$, black bonds: interdimer bonds $J_1$.
(Bottom) Magnetization curves at $T=0$ for $J_2=5$, $J_1=1$ and several uniform anisotropy parameters $\Delta_1=\Delta_2=\Delta$:
ED for the initial model ($N=16$; thin) vs the effective model (${\cal N}=8$, thick).
Blue lines: $\Delta=0.5$, green lines: $\Delta=0.1$, red lines: $\Delta=0$.
The solid lines correspond to ${\bf{h}}=(0,0,h)$ and the dashed lines to ${\bf{h}}=(h,0,0)$.}
\label{f01}
\end{center}
\end{figure}

An interesting consequence of the localized-magnon states for 2D Heisenberg FB systems 
is the prediction of a finite-temperature order-disorder phase transition
which should occur at low temperatures in a finite field region just below the saturation field
and is related to an ordering of the independent localized magnons \cite{Zhitomirsky2004,Richter2006,Derzhko2010b}.
The first prediction of such a transition \cite{Zhitomirsky2004} refers to the quantum kagome AFM 
where the compact localized states (located on hexagons) can be mapped onto the classical hard-hexagon problem \cite{Baxter1982}.
However, 
a specific problem for the kagome AFM is the existence of additional noncompact independent localized-magnon (LM) states \cite{Derzhko2007} 
which are not taken into account by the corresponding hard-hexagon problem, 
and, therefore the prediction of the phase transition is to some extent problematic. 
More promising is the fully frustrated square-lattice bilayer quantum Heisenberg AFM
(see Fig.~\ref{f01}, top), 
since for this model the compact localized states 
(located on the vertical interlayer bonds) 
are the only existing independent LM states.
For this model the ordering of localized magnons falls into the 2D Ising universality class \cite{Richter2006}.
Remarkably, for this system one can take into account an additional class of exact low-energy localized states 
(interacting/non-independent LM states)
which allows a comprehensive study of the ``high field -- low temperature'' phase diagram 
of the fully frustrated square-lattice bilayer spin-1/2 Heisenberg AFM \cite{Derzhko2010b}.
Very recently,
a specific feature of this model, 
namely the existence of local conservation laws, 
has been exploited to develop a new sign-problem-free Monte Carlo method \cite{Alet2016} 
that confirms the finite-temperature Ising transition in the square-lattice bilayer quantum spin system in a magnetic field
predicted in \cite{Richter2006,Derzhko2010b}. 

The question of the experimental observation of the particular LM physics in solid-state magnets 
is crucial for its relevance in material science.
Fortunately, there is a plethora of one-, two-, and three-dimensional frustrated spin models 
hosting independent localized magnons \cite{Schnack2001,Derzhko2006,Derzhko2015,Derzhko2007}.
Except the above mentioned kagome and bilayer AFMs, 
prominent examples are, 
e.g., the pyrochlore and the checkerboard AFMs  as well as the diamond spin chain.
On the other hand, 
the LM physics requires a certain fine-tuning of the Hamiltonian parameters and it takes place near the saturation field, 
that is often not accessible in experimental setups to measure thermodynamic quantities such as the specific heat. 
So far the most promising candidate was the natural mineral azurite \cite{Kikuchi2005},
which represents a one-dimensional frustrated diamond-chain Heisenberg AFM.
Although, its Hamiltonian parameters do not obey ideal FB conditions \cite{Jeschke2011,Derzhko2013,Richter2015}, 
it exhibits indeed a wide magnetization plateau ending in an almost perfect jump to saturation.

A very interesting candidate for 2D LM physics is Ba$_2$CoSi$_2$O$_6$Cl$_2$ \cite{Tanaka2014}.
Recently, Tanaka et al. \cite{Tanaka2014} have found that the magnetic Co$^{2+}$ ions of this compound 
can be described as a 2D fully frustrated square-lattice bilayer spin-1/2 AFM 
(see Fig.~\ref{f01}, top) 
with an antiferromagnetic vertical interlayer (intradimer) coupling $J_2$ 
dominating the nearest-neighbor intralayer couplings and the frustrating interlayer couplings.
Thus, the exchange pattern of Ba$_2$CoSi$_2$O$_6$Cl$_2$ perfectly fits to the LM (FB) conditions. 
Since the underlying magnetic model is 2D, 
the magnetic properties of Ba$_2$CoSi$_2$O$_6$Cl$_2$ are expected to be even more diverse than for the one-dimensional azurite.
The most spectacular experimental results are the magnetization curves $m(h)$ measured at 1.3~K for two field orientations until about 70~T,
that exceeds noticeably the saturation field.
The magnetization curve $m(h)$ exhibits a stepwise shape with a plateau at half of the saturation magnetization, 
irrespective of the field direction which is typical for the fully frustrated bilayer AFM \cite{Richter2006,Derzhko2010b}.
Although, the reported  magnetic properties of Ba$_2$CoSi$_2$O$_6$Cl$_2$ 
strongly resemble the theoretical predictions of \cite{Richter2006,Derzhko2010b} 
obtained for the fully frustrated isotropic Heisenberg bilayer AFM,
there are several important differences in the  appropriate spin model for Ba$_2$CoSi$_2$O$_6$Cl$_2$. 
First, 
the model is an anisotropic $XXZ$ spin model close to the $XY$ limit \cite{Tanaka2014}.
Second,
the $g$-factor for the field applied in the $XY$ plane is almost 2 times larger than the $g$-factor for the field applied along the $z$-axis, 
and, therefore, the corresponding saturation fields are quite different, 
namely ${\sf{H}}_{\rm{sat}}=41.0$~T (in $XY$ plane) and ${\sf{H}}_{\rm{sat}}=56.7$~T (along $z$-axis) \cite{Tanaka2014}.

\section{Effective theory for Ba$_2$CoSi$_2$O$_6$Cl$_2$}

In the present paper we develop a theory for Ba$_2$CoSi$_2$O$_6$Cl$_2$ 
that is based on the LM picture for the bilayer model, 
see \cite{Richter2006,Derzhko2010b}. 
Using this theory we want to describe the reported experimental magnetization curves 
and to propose new experiments to detect specific features of the LM physics, 
such as an extra low-temperature singularity in the specific heat and a magnetic-field driven order-disorder phase transition.
Let us first mention that the LM scenario also holds 
for $XXZ$ models with magnetic fields along $z$-direction \cite{Schnack2001,Richter2004},
whereas the case when the Zeeman term does not commute with the $XXZ$ interactions 
(e.g., with a magnetic field along $x$-direction) 
was not studied so far.

Extending the theory of \cite{Richter2006,Derzhko2010b} 
we consider an anisotropic spin-1/2 $XXZ$  square-lattice bilayer AFM of $N=2{\cal{N}}$ sites as shown in Fig.~\ref{f01}
and introduce different values of the $g$-factor for the field directed along $x$-axis and along $z$-axis
\begin{eqnarray}
\label{01}
H=\sum_{\langle i j\rangle}J_{ij} \left(s^x_i s^x_j + s^y_i s^y_j + \Delta_{ij} s^z_i s^z_j\right) -\sum_{i=1}^N
{\bf{h}}\cdot{\bf{s}}_i,
\end{eqnarray}
where $J_{ij}$ and $\Delta_{ij}$ acquire either the values $J_2$ and $\Delta_2$ (vertical dimer bonds) or $J_1$ and $\Delta_1$ (interdimer bonds), 
cf. Fig.~\ref{f01}.
Note that the equality of all interdimer bonds corresponds just to the ideal FB geometry. 
Moreover, a sufficiently large value of $J_2$ is needed to establish LM physics 
(e.g., $J_2>4J_1$ for $\Delta_1=\Delta_2=1$ \cite{Richter2006,Derzhko2010b}).
In correspondence to \cite{Tanaka2014} 
we consider two particular orientations of the field:
${\bf{h}}=(0,0,h^z)$ and ${\bf{h}}=(h^x,0,0)$
and corresponding $g$-factors: 
$g^z$ and $g^x$,
i.e.,
$h^z=g^z\mu_{\rm{B}}{\sf{H}}$ and $h^x=g^x\mu_{\rm{B}}{\sf{H}}$,
where 
$\mu_{\rm{B}}\approx 0.671\,71$~K/T is the Bohr magneton
and 
${\sf{H}}$ is the value (measured in Tesla) of the applied magnetic field.
According to \cite{Tanaka2014},
the $g$-factors for Ba$_2$CoSi$_2$O$_6$Cl$_2$ are $g^z=2.0\pm 0.1$ and $g^x=3.86$.

Next we elaborate an effective low-energy theory for the model (\ref{01}) at high magnetic fields 
by using the strong-coupling approach for both cases of the field direction.
For that we assume 
that the main part of the Hamiltonian $H_{\rm{main}}$ consists only of the vertical bonds $J_2$ and the Zeeman term at the field $h_0$, 
where the two eigenstates, $\vert u\rangle$ and $\vert d\rangle$, of the spin dimer are degenerate. 
Hence, at $h_0$ a magnetization jump to saturation (to almost saturation) takes place for $z$-aligned ($x$-aligned) field.
The remaining terms in (\ref{01}) are treated as the perturbation $V=H-H_{\rm{main}}$.
The effective Hamiltonian is obtained by standard perturbation theory \cite{Honecker2001,Mila2011,Fulde1993}
$H_{\rm{eff}}=PHP+ \ldots$,
where $P=\vert\varphi_0\rangle\langle\varphi_0\vert$ is the projector onto the ground-state manifold of $H_{\rm{main}}$ 
consisting of $2^{\cal{N}}$ states of the ${\cal{N}}$ dimers. 
In addition, 
we use the (pseudo)spin-1/2 operators 
$T^z=(\vert u\rangle\langle u\vert - \vert d\rangle\langle d\vert)/2$,
$T^+=\vert u\rangle\langle d\vert$,
$T^-=\vert d\rangle\langle u\vert$,
attached to each vertical dimer bond
to represent $H_{\rm{eff}}$ in an easy recognizable form.
After some straightforward manipulations (see Appendix) we obtain the effective model  
\begin{eqnarray}
\label{02}
H_{\rm{eff}}
=
{\sf C}
-{\sf{h}}\sum_{m=1}^{\cal{N}} T^z_{m}
+{\sf{J}}\sum_{\langle mn\rangle} T^z_{m}T^z_{n}, 
\end{eqnarray}
that corresponds to the square-lattice spin-1/2 antiferromagnetic Ising model.
The parameters  ${\sf C}$, ${\sf{h}}$ and ${\sf{J}}$  
are functions of the Hamiltonian parameters $J_2$, $\Delta_2$, $J_1$, $\Delta_1$, and $h$.
For the  $z$-directed field these parameters  are given by simple formulas:
${\sf C}=(-h/2-J_2/4+\Delta_{1}J_1/2){\cal{N}}$,
${\sf{h}}=h-(1+\Delta_2)J_2/2-2\Delta_{1}J_1$,
and
${\sf J}=\Delta_{1}J_1$.
In case of an  $x$-directed field,
these expressions are more complicated (see Appendix).
Since the  field $\sf{h}$ in (\ref{02}) 
is a function of $J_i$, $\Delta_i$ and, of course, of the applied magnetic field ${\sf{H}}$,
for a certain value of ${\sf{H}}$ the resulting effective field ${\sf{h}}$ is zero, 
i.e., the effective model (\ref{02}) is the exactly solvable zero-field square-lattice Ising model \cite{Onsager1944}.
For ${\sf{h}} \ne 0$,
there are high-precision numerical results for the phase diagram in the ``field -- temperature'' plane, 
see, e.g., \cite{Muller-Hartmann1977,Wu1990,Wang1997,Penney2003,Viana2009}.
The ordered phase of the effective Ising model 
corresponds to a Wigner-crystal ordering of localized magnons in the initial quantum model.

\section{Field-driven phase transition in Ba$_2$CoSi$_2$O$_6$Cl$_2$}

In a first step we check the quality of the elaborated effective description.
For that we perform full exact diagonalization (ED) \cite{Richter2010} 
for both, the initial and the effective models on finite lattices imposing periodic boundary conditions.
For the initial (quantum) model (\ref{01}) that is the $N=16=2\times 8$ bilayer 
and 
for the effective (classical) model (\ref{02}) that it is the ${\cal{N}}=8$ square lattice. 
In Fig.~\ref{f01}, bottom, we compare the ground-state magnetization curves taking as a representative the parameter set:
$J_2=5$, $J_1=1$, $\Delta_2=\Delta_1=\Delta$ with $\Delta=0.5,\, 0.1,\, 0$
and $z$-aligned  (solid curves) as well as  $x$-aligned (dashed curves) magnetic fields. 
For the $z$-aligned field the curves of both models practically coincide,
whereas for the $x$-aligned field the agreement is still very good, 
in particular at the one-half plateau, where the LM-crystal state is an exact eigenstate of (\ref{01}).

Now we turn to the specific situation of Ba$_2$CoSi$_2$O$_6$Cl$_2$.
The strong intradimer interaction parameters, 
estimated as $J_2=110$~K, $\Delta_2=0.149$ \cite{Tanaka2014}, 
dictate the low-temperature magnetization process.
As mentioned above,  all interdimer bonds are practically equal,
i.e., the ideal FB condition holds in good approximation, 
and their values given in \cite{Tanaka2014} are $4J_1=23.7$~K and $\Delta_1 = 0.56$.
We also recall the $g$-factors $g^z=2.0$ and $g^x=3.86$ \cite{Tanaka2014}.
Putting all pieces together and measuring the applied magnetic field ${\sf{H}}$ in Tesla, 
we arrive at the following estimates for the parameters of $H_{\rm{eff}}$, Eq.~({\ref{02}):
for the $z$-aligned field ${\sf{H}}$ we get
${\sf C} \approx (-0.67{\sf{H}}-25.84){\cal{N}}$~K,
${\sf h}\approx -(1.34{\sf{H}}-69.85)$~K,
${\sf J} \approx 3.33$~K. 
In case of an $x$-aligned field ${\sf{H}}$ the effective parameters are given by more complicated formulas 
(see Appendix).
However, in the field range from 30~T to 45~T, relevant for the localized magnon scenario, 
these parameters  vary almost linearly with ${\sf{H}}$ as follows:
${\sf C} \approx  -53.70 \ldots -72.45$~K,
${\sf{h}} \approx -16.24 \ldots 21.01$~K,
and 
${\sf{J}} \approx 5.43 \ldots 5.70$~K.
Note, however, that for the results discussed below we use the full formulas of the effective parameters 
given in Appendix.

\begin{figure}
\begin{center}
\includegraphics[clip=on,width=60mm,angle=0]{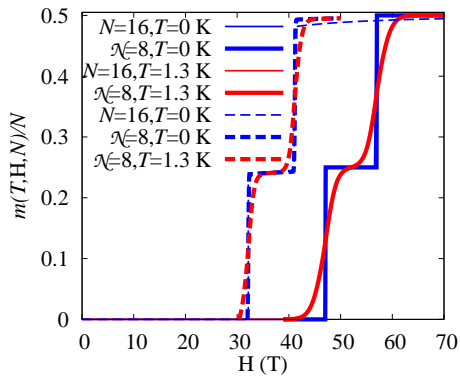}
\caption{Magnetization curves for the model parameters of Ba$_2$CoSi$_2$O$_6$Cl$_2$
($J_2=110$~K, $\Delta_2=0.149$, $4J_1=23.7$~K, $\Delta_1=0.56$) \cite{Tanaka2014} 
at $T=0$~K (blue) and $T=1.3$~K (red).
ED data for the initial model ($N=16$, thin lines) 
are compared with corresponding ones for the effective model (${\cal{N}}=8$, thick lines). 
Note that corresponding thin and thick lines widely coincide. 
Solid lines represent data for ${\bf{h}}=(0,0,h^z)$, $g^z=2.0$ 
and  
dashed lines for ${\bf{h}}=(h^x,0,0)$, $g^x=3.86$, 
where ${\sf{H}}=h^\alpha/(g^\alpha\mu_{\rm{B}})$, $\alpha=z,x$.}
\label{f02}
\end{center}
\end{figure}

In Fig.~\ref{f02} we report ED results for magnetization curves for both field directions.
These curves fit well to the experimental ones, see Fig.~3 in \cite{Tanaka2014}
(note, however, that the experimental data are calibrated differently).
We also mention that $T=1.3$ K used to calculate the isothermal $m({\sf{H}})$ curve 
cannot be identified as that temperature quoted in Fig.~3 of \cite{Tanaka2014}, 
because in \cite{Tanaka2014} a pulsed magnetic field was used to measure $m({\sf{H}})$, 
i.e., the measurement process is rather adiabatic than isothermal.
The magnetization jumps in Fig.~\ref{f02} are located at the critical fields
${\sf{H}}_c \approx 47.04$~T and ${\sf{H}}_{\rm{sat}} \approx 56.94$~T for the $z$-directed field 
and
${\sf{H}}_c \approx 32.16$~T and ${\sf{H}}_{\rm{sat}} \approx 41.14$~T for the $x$-directed field.
These values agree well with the corresponding experimental data \cite{Tanaka2014}
${\sf{H}}_c=46.8$~T, ${\sf{H}}_{\rm{sat}}=56.7$~T
and
${\sf{H}}_c=32.0$~T, ${\sf{H}}_{\rm{sat}}=41.0$~T.
Thus, we may conclude 
that our theoretical approach provides an excellent description of the measured magnetization process of Ba$_2$CoSi$_2$O$_6$Cl$_2$.

Next we analyze the specific heat $c$ in the plateau region ${\sf{H}}_c < {\sf{H}} < {\sf{H}}_{\rm{sat}}$ 
predicted by theory, but not measured yet.
For the $z$-aligned field $ {\sf H} \approx 51.99$~T and for the $x$-aligned field $ {\sf H}\approx 36.61$~T 
the effective field $ {\sf h}$ in Eq.~(\ref{02}) vanishes, 
and the corresponding effective coupling constants ${\sf J}$ yield the transition temperatures  
$T_c\approx 1.89$~K ($z$-aligned field) 
and 
$T_c\approx 3.17$~K ($x$-aligned field). 

\begin{figure}
\begin{center}
\includegraphics[clip=on,width=60mm,angle=0]{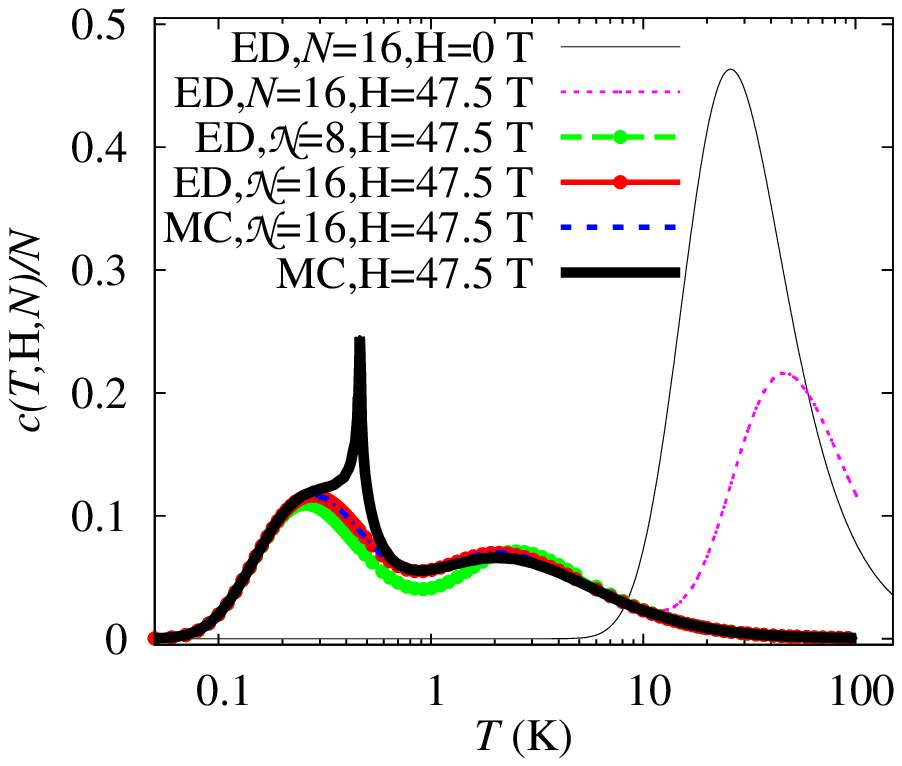}
\includegraphics[clip=on,width=60mm,angle=0]{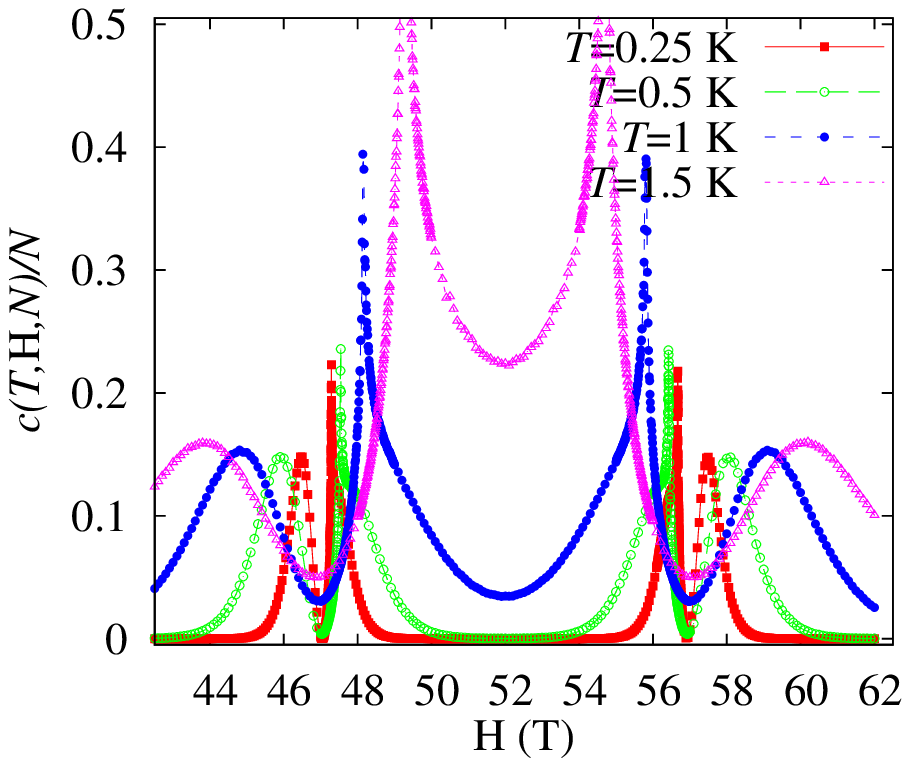}
\caption{(Top) ED (initial and effective model with $N=16$ and ${\cal{N}}=8$) and MC data (effective model with ${\cal{N}}$ up to $256 \times 256$) 
for the temperature dependence of the specific heat 
for the model parameters of Ba$_2$CoSi$_2$O$_6$Cl$_2$
(see text)
at ${\sf{H}}=47.5$~T along the $z$-axis ($T_c\approx 0.46$~K).
(Bottom) MC data (effective model with ${\cal{N}}$ up to $256\times 256$) 
for the specific heat 
for the model parameters of Ba$_2$CoSi$_2$O$_6$Cl$_2$
(see text)
at $T=0.25,\,0.5,\,1,\,1.5$~K for the field applied along the $z$-axis.}
\label{f03}
\end{center}
\end{figure}

\begin{figure}
\begin{center}
\includegraphics[clip=on,width=60mm,angle=0]{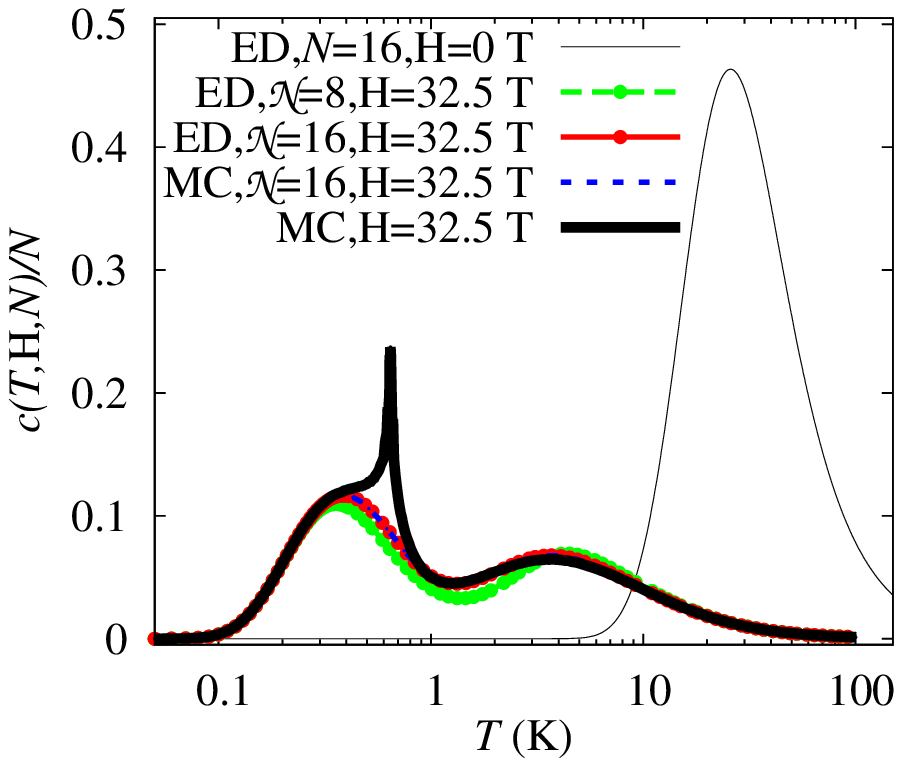}
\includegraphics[clip=on,width=60mm,angle=0]{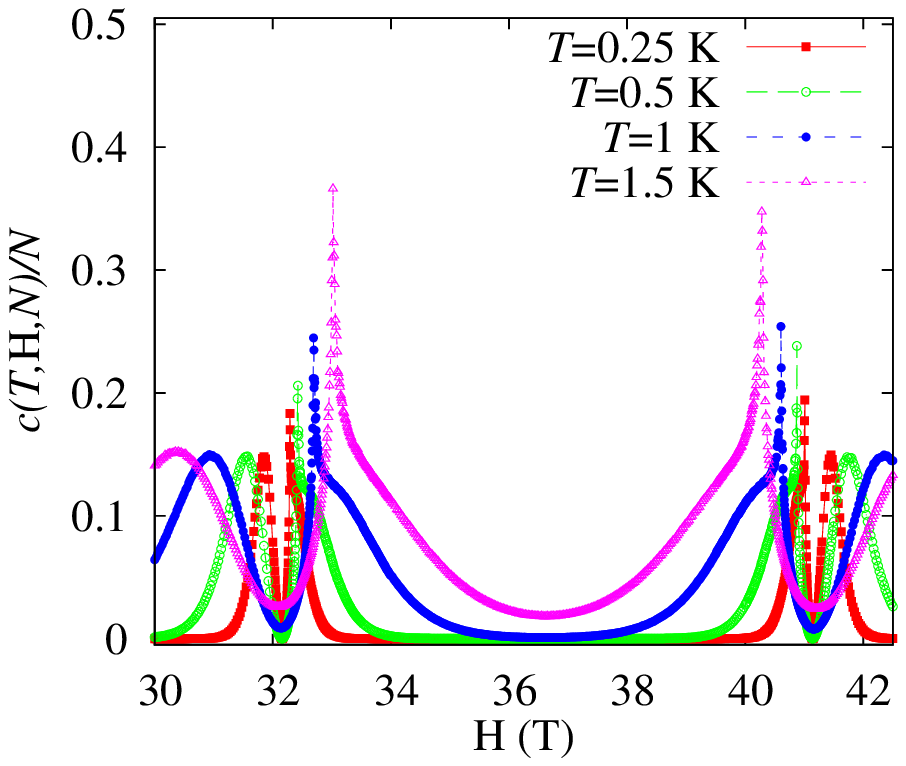}
\caption{(Top) 
ED (initial and effective model with $N=16$ and ${\cal{N}}=8$) and MC data (effective model with ${\cal{N}}$ up to $256\times 256$)
for the specific heat for the model parameters of Ba$_2$CoSi$_2$O$_6$Cl$_2$ (see text
of the main paper)
at ${\sf{H}}=32.5$~T along $x$-axis ($T_c\approx 0.64$~K).
(Bottom) 
MC data (effective model with ${\cal{N}}$ up to $256 \times 256 $) 
for the specific heat for the model parameters of Ba$_2$CoSi$_2$O$_6$Cl$_2$ (see text
of the main paper)
at $T=0.25,\,0.5,\,1,\,1.5$~K for the field applied along the $x$-axis.}
\label{Sf08}
\end{center}
\end{figure}

In the upper panel of Fig.~\ref{f03} we show ED data for the specific heat $c(T)$ at ${\sf{H}}=47.5$~T along $z$-axis 
for the effective model (${\cal N}=8$ and $16$) and for the initial model ($N=16$) 
as well as classical Monte Carlo (MC) data for the effective model of ${\cal N}={\cal{L}}^2$ sites with ${\cal{L}}$ up to 256. 
For large enough system size the Ising-Onsager logarithmic singularity at $T_c\approx 0.46$~K is evident. 
In contrast, no singularity is present for $c(T)$ at ${\sf{H}}=0$.
It is also evident that the finite-size data for the effective and the full initial model coincide up to about $T\approx 10$~K 
[cf. green (${\cal{N}}=8$) and magenta ($N=16$) lines in the upper panel of Fig.~\ref{f03}], 
i.e., far beyond $T_c\approx 0.46$~K.
Except the singularity at $T_c$ we see clear signatures of a separation of energy scales indicated by two maxima in the effective model, 
and even three maxima in the full initial model, 
where the highest-temperature one is related to the strength of $J_2$. 
This scale is not present in the effective model, where the interaction strength ${\sf J}\propto J_1$.
The intermediate-temperature maximum is related to ${\sf J}$, 
the lowest-temperature one corresponds to the energy scale 
set by the degenerated manifold of states being ground states at ${\sf{H}}_{\rm{sat}}$ and ${\sf{H}}_c$.
The position of the lowest-temperature maximum depends on the value of ${\sf{H}}$, ${\sf{H}}_{c}<{\sf H}<{\sf{H}}_{\rm{sat}}$,
and, it moves to $T=0$ as ${\sf H}\to {\sf{H}}_{\rm{sat}}$ as well as ${\sf H}\to {\sf{H}}_{c}$.
Another way to detect the phase transition is to fix $T$ and use ${\sf{H}}$ as the driving parameter. 
A corresponding plot is shown in the lower panel of Fig.~\ref{f03},  
where MC data of $c({\sf{H}})$ for the $z$-aligned field at fixed $T=0.25,\,0.5,\,1,\,1.5$~K are presented. 
Note that a very similar behavior of $c(T)$ and $c({\sf H})$ is found for the $x$-aligned field, see Fig.~\ref{Sf08}.
Thus corresponding measurements on Ba$_2$CoSi$_2$O$_6$Cl$_2$ are highly desirable to verify our predictions.

\begin{figure}
\begin{center}
\includegraphics[clip=on,width=60mm,angle=0]{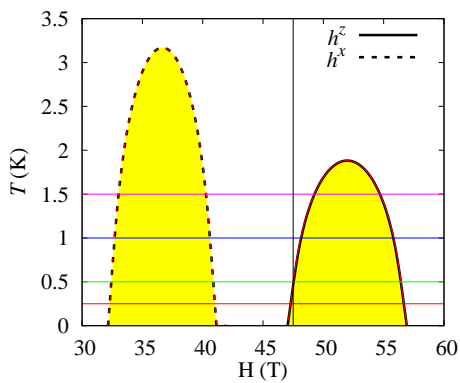}
\caption{Theoretically predicted phase diagram for Ba$_2$CoSi$_2$O$_6$Cl$_2$ in the ``field -- temperature'' plane
for ${\bf{h}}=(0,0,h^z)$ (solid) and ${\bf{h}}=(h^x,0,0)$ (dashed).
The presented curves were obtained from the results of \cite{Wu1990,Wang1997}.
The horizontal (vertical) lines correspond to the lines of same color in
Fig.~\ref{f03}, bottom (Fig.~\ref{f03}, top).}  
\label{f04}
\end{center}
\end{figure}

The main result summarizing our findings 
is the phase diagram for Ba$_2$CoSi$_2$O$_6$Cl$_2$ in the ``field -- temperature'' plane 
shown in Fig.~\ref{f04}.
From this phase diagram one concludes 
that applying an $x$-aligned ($z$-aligned) field ${\sf{H}} > 32.16$~T (${\sf{H}} > 47.04$~T) to Ba$_2$CoSi$_2$O$_6$Cl$_2$ 
one should observe an ordering of localized magnons (magnon Wigner crystal), 
where the corresponding phase transition belongs to the 2D Ising model universality class.
This phase transition can be detected by measuring the specific heat $c$, 
which exhibits a logarithmic singularity 
in its temperature dependence at $T_{\rm trans}({\sf{H}})$ 
or 
in its field dependence at ${\sf{H}}_{\rm trans}(T)$, 
both given by the transition lines shown in Fig.~\ref{f04}.

\section{Conclusions}

To conclude, we demonstrate 
that Ba$_2$CoSi$_2$O$_6$Cl$_2$ is a promising candidate to realize FB physics in a highly frustrated quantum magnet.
Based on the concept of localized magnons we provide a theory 
to describe experimental data in high magnetic fields \cite{Tanaka2014}.
The most important result of our theory is the prediction of a phase transition related to an ordering of the localized magnons.
This phase transition occurs in high magnetic fields ${\sf H} > 32.16$~T 
and can be driven either by temperature or by magnetic field.
To detect this transition in experiment  
we propose low-temperature measurements of the specific heat $c$ at ${\sf H} > 32.16$~T to find the characteristic singularity in $c$.

\section*{Acknowledgments}

The present study was supported by the Deutsche Forschungsgemeinschaft (project RI615/21-2).
J.~R. and O.~D. are grateful 
to H.~Tanaka 
for fruitful discussions during the International Workshop {\it Flatband Networks in Condensed Matter and Photonics} (August 28 -- September 1, 2017)
and 
to the Center for Theoretical Physics of Complex Systems of the Institute for Basic Science (Daejeon, Korea) 
for financial support and hospitality.

\section*{Appendix: Derivation of the effective model (\ref{02})}
\renewcommand{\theequation}{A\arabic{equation}}
\setcounter{equation}{0}

In this appendix,
we present some technical details of calculations leading to the effective theories,
which are reported in the main text.

Let us consider the spin-1/2 Heisenberg antiferromagnet
for a more general square-lattice bilayer of $N=2{\cal{N}}$ sites
\begin{eqnarray}
\label{S01}
H=\sum_{\langle i j\rangle}J_{ij} \left(s^x_i s^x_j + s^y_i s^y_j + \Delta_{ij} s^z_i s^z_j\right)
-\sum_{i=1}^N {\bf{h}}\cdot{\bf{s}}_i.
\end{eqnarray}
The first sum in Eq.~(\ref{S01}) runs over all bonds of the lattice 
and hence $J_{ij}$ acquires either the value $J_2$ (dimer bonds) or $J_{11}$, $J_{12}$, $J_{21}$, $J_{22}$ (all other bonds), 
see Fig.~\ref{Sf06}.
The important limiting case considered in the main text corresponds to the relation $J_{11}=J_{12}=J_{21}=J_{22}=J_1$.
This ideal frustration case for the isotropic Heisenberg case 
(i.e., $\Delta_{ij}=1$) was considered in Refs.~\cite{Richter2006,Derzhko2010b}.
The crucial difference to the previous studies \cite{Richter2006,Derzhko2010b}
is the anisotropy of the exchange interaction controlled by the parameter $0\le \Delta_{ij}\le 1$
($\Delta_2=0.149$ and $\Delta_1=0.56$ are the values relevant for Ba$_2$CoSi$_2$O$_6$Cl$_2$, see Ref.~\cite{Tanaka2014}).
Moreover, the external magnetic field ${\bf{h}}=(h^x,h^y,h^z)$ may have, in principal, an arbitrary orientation. 
For the sake of simplicity, 
we consider in what follows two particular orientations of the field:
${\bf{h}}=(0,0,h^z)$ (directed along $z$-axis)
and
${\bf{h}}=(h^x,0,0)$ (directed along $x$-axis).
While in the former case we start with the Hamiltonian given in Eq.~(\ref{S01}) with ${\bf{h}}=(0,0,h^z)$,
that is,
\begin{eqnarray}
\label{S03}
H=\sum_{\langle i j\rangle}J_{ij} \left(s^x_i s^x_j + s^y_i s^y_j + \Delta_{ij} s^z_i s^z_j\right) -h^z\sum_{i=1}^N s^z_i,
\end{eqnarray}
in the latter case it is convenient to rotate the spin axes arriving at the Hamiltonian
\begin{eqnarray}
\label{S04}
H=\sum_{\langle i j\rangle}J_{ij} \left[{\bf{s}}_i\cdot{\bf{s}}_j
+\left(\Delta_{ij}-1\right)s^x_i s^x_j\right] - h^x\sum_{i=1}^N s^z_i.
\end{eqnarray}
We have also to take into account the difference in $g$-factors for the field applied along $z$-axis and $x$-axis.
For this we set
\begin{eqnarray}
\label{S05}
h^z=g^z\mu_{\rm{B}}{\sf{H}},
\;\;\;\;\;
h^x=g^x\mu_{\rm{B}}{\sf{H}},
\end{eqnarray}
where 
$g^z$ and $g^x$ are the values of $g$-factor,
$\mu_{\rm{B}}\approx 0.671\,71$~K/T is  the Bohr magneton,
and ${\sf{H}}$ is the value (measured in Tesla) of applied magnetic field in experiments.
According to Ref.~\cite{Tanaka2014},
the $g$-factors for Ba$_2$CoSi$_2$O$_6$Cl$_2$ were determined as $g^z=2.0\pm 0.1$ and $g^x=3.86$.

\begin{figure}
\begin{center}
\includegraphics[clip=on,width=60mm,angle=0]{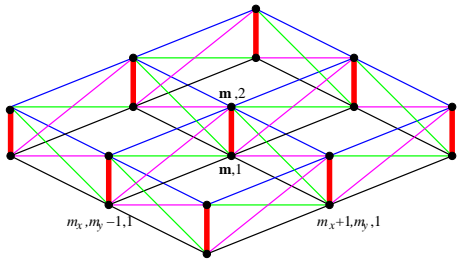}
\caption{Frustrated square-lattice bilayer.
The sites are numerated by the cell index $m_x,m_y$ and the site index within a cell $i=1,2$.
The exchange interaction for the vertical bond (thick red bonds) is $J_2$,
whereas the exchange interaction for the bond connecting 
the sites $m_x,m_y,i$ and $m_x+1,m_y,j$
or 
the sites $m_x,m_y,i$ and $m_x,m_y+1,j$
(thin bonds)
is $J_{ij}$.}
\label{Sf06}
\end{center}
\end{figure}

In general, 
the concept of localized magnons can be introduced in the case when the $z$-component of the total spin commutes with the Hamiltonian
[as for the Hamiltonian (\ref{S03})]:
then the eigenstates of the Hamiltonian can be examined in the subspaces with different values of $S^z=\sum_{i=1}^N s_i^z$ separately. 
However, even for the Hamiltonian  (\ref{S04}),
the LM-crystal state
(but not the ground state in the extremely-strong-field limit)
is still an exact eigenstate.

Now we elaborate an effective low-energy theory of the introduced model 
[Eq.~(\ref{S01}), Fig.~\ref{Sf06}]
at high magnetic fields.
To achieve this goal, we use the strong-coupling approach.
We consider separately two cases:
the field directed along $z$-axis [see Eq.~(\ref{S03})]
and
the field directed along $x$-axis [see Eq.~(\ref{S04})]. 
Within the strong-coupling approach,
we assume that the main part of the Hamiltonian $H_{\rm{main}}$ consists of the spins on vertical bond only at the ``bare'' saturation field $h_0$.
At this field, $h=h_0$, the energy of two states $\vert u\rangle$ and $\vert d\rangle$ of the two-spin system on the vertical bond coincides.
The rest terms in the Hamiltonian (\ref{S03}) or (\ref{S04}) are treated as the perturbation $V=H-H_{\rm{main}}$.
The wanted effective Hamiltonian follows from the perturbation-theory formula \cite{Honecker2001,Mila2011,Fulde1993}
\begin{eqnarray}
\label{S06}
H_{\rm{eff}}=PHP+PV\sum_{\alpha\ne 0}\frac{\vert\varphi_\alpha\rangle\langle\varphi_\alpha\vert}{\varepsilon_0-\varepsilon_\alpha}VP+\ldots.
\end{eqnarray}
Here $P=\vert\varphi_0\rangle\langle\varphi_0\vert$ is the projector onto the ground-state manifold of $H_{\rm{main}}$ 
consisting of $2^{\cal{N}}$ states,
and $\varepsilon_0$ and $\varepsilon_\alpha$ are the energies of the ground state and excited states of $H_{\rm{main}}$, respectively.
In what follows we restrict ourselves to the first term in the r.h.s. of Eq.~(\ref{S06}) only.
In addition, 
we use the (pseudo)spin-1/2 operators 
$T^z=(\vert u\rangle\langle u\vert - \vert d\rangle\langle d\vert)/2$,
$T^+=\vert u\rangle\langle d\vert$,
$T^-=\vert d\rangle\langle u\vert$,
attached to each vertical bond
to represent the effective Hamiltonian (\ref{S06}) in an easy recognizable form.

We begin with the case of $z$-directed field, see Eq.~(\ref{S03}).
Two relevant states at each dimer,
$\vert u\rangle=\vert\uparrow\uparrow\rangle$ 
and 
$\vert d\rangle =(\vert\uparrow\downarrow\rangle-\vert\downarrow\uparrow\rangle)/\sqrt{2}$,
have identical energies at $h_0=(1+\Delta_2)J_2/2$.
The effective Hamiltonian is given by the formula:
\begin{eqnarray}
\label{S07}
H_{\rm{eff}}
=
\left(-\frac{h}{2}-\frac{J_2}{4}+\frac{J_{\Delta}}{2}\right){\cal{N}}
-{\sf{h}}\sum_{m=1}^{\cal{N}} T^z_{m}
\nonumber\\
+
\sum_{\langle mn\rangle}
\left[{\sf{J}}^zT^z_{m}T^z_{n}
+
{\sf{J}}\left(T^x_{m}T^x_{n} + T^y_{m}T^y_{n}\right)
\right].
\end{eqnarray}
Here the parameters of the effective Hamiltonian are defined as follows:
\begin{eqnarray}
\label{S08}
J_{\Delta}=\frac{\Delta_{11}J_{11}+\Delta_{12}J_{12}+\Delta_{21}J_{21}+\Delta_{22}J_{22}}{4},
\nonumber\\
{\sf{h}}=h-h_1,
\;\;\;
h_1=\frac{1+\Delta_2}{2}J_2+2J_{\Delta},
\nonumber\\
{\sf{J}}^z=J_{\Delta},
\nonumber\\
{\sf{J}}=\frac{J_{11}-J_{12}-J_{21}+J_{22}}{2}.
\end{eqnarray}
Clearly,
we have arrived at the square-lattice spin-1/2 $XXZ$ Heisenberg model in a $z$-aligned magnetic field.
This model is free of frustration and can be studied,
for example,
employing quantum Monte Carlo method.

It is worth noting that in the case $J_{11}=J_{12}=J_{21}=J_{22}=J_1$,
Eqs.~(\ref{S07}), (\ref{S08}) correspond to the square-lattice spin-1/2 antiferromagnetic Ising model in a field,
\begin{eqnarray}
\label{S09}
H_{\rm{eff}}
=
\left(-\frac{h}{2}-\frac{J_2}{4}+\frac{J_{\Delta}}{2}\right){\cal{N}}
\nonumber\\
-\left(h-\frac{1+\Delta_2}{2}J_2-2J_{\Delta}\right)\sum_{m=1}^{\cal{N}} T^z_{m}
+
J_{\Delta} \sum_{\langle mn\rangle} T^z_{m}T^z_{n},
\nonumber\\
J_{\Delta}=\frac{\Delta_{11}+\Delta_{12}+\Delta_{21}+\Delta_{22}}{4}J_1,
\end{eqnarray}
see also Eq.~(\ref{S01}).
The obtained result is in agreement with previous ones \cite{Derzhko2010b}.
Really,
Eq.~(\ref{S09}) at $\Delta_2=\Delta_{11}=\ldots=\Delta_{22}=1$ 
yields the lattice-gas model with finite repulsion given in Eq.~(6.7) of Ref.~\cite{Derzhko2010b}.

Next,
we consider the case of $x$-directed field, see Eq.~(\ref{S04}).
Now the two relevant states at each dimer and their energies are \cite{Richter2015}:
\begin{eqnarray}
\label{S10} 
\vert u\rangle=a\vert\uparrow\uparrow\rangle +b\vert\downarrow\downarrow\rangle,
\nonumber\\
a=\frac{1}{C}
\left[
h+\sqrt{\frac{(1-\Delta_2)^2}{16}J_2^2+h^2}
\right],
\;
b=\frac{1}{C}\frac{1-\Delta_2}{4}J_2,
\nonumber\\
C=\sqrt{2}
\sqrt{\frac{(1-\Delta_2)^2}{16}J_2^2+ h\sqrt{\frac{(1-\Delta_2)^2}{16}J_2^2+h^2} +h^2 },
\nonumber\\
\epsilon_u=\frac{J_2-\sqrt{(1-\Delta_2)^2J_2^2+16h^2}}{4}
\end{eqnarray}
and
\begin{eqnarray}
\label{S11} 
\vert d\rangle=\frac{1}{\sqrt{2}}\left(\vert\uparrow\downarrow\rangle -\vert\downarrow\uparrow\rangle\right),
\nonumber\\
\epsilon_d=-\frac{(2+\Delta_2)J_2}{4}.
\end{eqnarray}
Furthermore, $h_0=\sqrt{(1+\Delta_2)/2} J_2$.
The effective Hamiltonian has the following form:
\begin{eqnarray}
\label{S12}
H_{\rm{eff}}
=
-\frac{\sqrt{(1-\Delta_2)^2J_2^2+16h^2}+(1+\Delta_2)J_2}{8}{\cal{N}}
\nonumber\\
-{\sf{h}}\sum_{m=1}^{\cal{N}} T^z_{m}
+
\sum_{\langle mn\rangle}
\left[{\sf{J}}\left(\frac{1}{2}+T^z_{m}\right)\left(\frac{1}{2}+T^z_{n}\right)
\right.
\nonumber\\
\left.
+{\sf{J}}^{+-}\left(T^+_{m}T^-_{n} + T^-_{m}T^+_{n}\right)
+{\sf{J}}^{++}\left(T^+_{m}T^+_{n} + T^-_{m}T^-_{n}\right)
\right],
\end{eqnarray}
where
\begin{eqnarray}
\label{S13}
{\sf{h}}=\frac{\sqrt{(1-\Delta_2)^2J_2^2+16h^2} - (3+\Delta_2)J_2}{4},
\nonumber\\
{\sf{J}}=\left(a^2-b^2\right)^2\frac{J_{11}+J_{12}+J_{21}+J_{22}}{4},
\nonumber\\
{\sf{J}}^{+-}
=\frac{{\cal{J}}^{+}_{11}-{\cal{J}}^{+}_{12}-{\cal{J}}^{+}_{21}+{\cal{J}}^{+}_{22}}{4},
\nonumber\\
{\sf{J}}^{++}
=\frac{{\cal{J}}^{-}_{11}-{\cal{J}}^{-}_{12}-{\cal{J}}^{-}_{21}+{\cal{J}}^{-}_{22}}{4},
\nonumber\\
{\cal{J}}^{\pm}_{ij}=\frac{(a-b)^2\Delta_{ij}\pm(a+b)^2}{2}J_{ij}.
\end{eqnarray}
Thus, $H_{\rm{eff}}$ represents  the square-lattice spin-1/2 $XYZ$ Heisenberg model in a $z$-aligned magnetic field.
In the case $J_{11}=J_{12}=J_{21}=J_{22}=J_1$ and $\Delta_{11}=\Delta_{12}=\Delta_{21}=\Delta_{22}=\Delta_1$,
important simplifications occur: 
${\sf{J}}=(a^2-b^2)^2 J_1$,
${\sf{J}}^{+-}={\sf{J}}^{++}=0$,
and Eqs.~(\ref{S12}), (\ref{S13}) transform into the Hamiltonian of the square-lattice spin-1/2 antiferromagnetic Ising model in a field,
\begin{eqnarray}
\label{S14}
H_{\rm{eff}}
=
-\frac{\sqrt{(1-\Delta_2)^2J_2^2+16h^2}+(1+\Delta_2)J_2-4{\sf{J}}}{8}{\cal{N}}
\nonumber\\
-\left({\sf{h}}-2{\sf{J}}\right)\sum_{m=1}^{\cal{N}} T^z_{m}
+{\sf{J}}\sum_{\langle mn\rangle} T^z_{m}T^z_{n}, 
\nonumber\\
{\sf{J}}=(a^2-b^2)^2 J_1,
\nonumber\\
{\sf{h}}=\frac{\sqrt{(1-\Delta_2)^2J_2^2+16h^2} - (3+\Delta_2)J_2}{4}
\end{eqnarray}
with $a$ and $b$ given in Eq.~(\ref{S10}).
If in addition $\Delta_2=1$, 
we have $a=1$, $b=0$, ${\sf{h}}=h-J_2$, ${\sf{J}}=J_1$, 
and Eq.~(\ref{S14}) transforms into Eq.~(\ref{S09}) at $\Delta_2=\Delta_{11}=\ldots=\Delta_{22}=1$.

\end{document}